\documentclass[showpacs,prb,twocolumn]{revtex4}
\usepackage{graphicx}
\usepackage{enumerate}
\usepackage{hyperref}
\usepackage{textcomp}
\usepackage{amsmath}
\usepackage{amssymb}
\usepackage{pgf}
\usepackage{diagbox}
\usepackage{booktabs}
\usepackage{bbold}
\usepackage{subfigure}
\usepackage{epstopdf}
\usepackage{color}
\usepackage{soul}

\newcommand{\Tr}{{\mathrm{Tr}}}

\newcommand{\nep}{\textrm{e}}

\newcommand{\LZ}{\mathrm{\scriptscriptstyle LZ}}

\newcommand{\Relax}{\mathrm{\scriptscriptstyle R}}
\newcommand{\Deph}{{\scriptscriptstyle \varphi}}
\newcommand{\Deco}{\mathrm{\scriptscriptstyle D}}

\newcommand{\opbdag}[1]{{\hat{b}^{\dagger}}_{#1}}
\newcommand{\opb}[1]{{\hat{b}^{\phantom \dagger}}_{#1}}

\newcommand{\gs}{\mathrm{gs}}

\newcommand{\PauliSigma}{\hat{\sigma}}

\newcommand{\hc}{\mathrm{H.c.}}
\newcommand{\Texp}{\mathcal{T} \mathrm{exp}}
\newcommand{\id}{\mathbb{1}}

\begin{document}

\title{Dissipative Landau-Zener problem and thermally assisted Quantum Annealing}

\author{Luca Arceci$^{1}$, Simone Barbarino$^{1}$, Rosario Fazio$^{3,4}$, Giuseppe E. Santoro$^{1,2,3}$}

\affiliation{
$^1$ SISSA, Via Bonomea 265, I-34136 Trieste, Italy\\
$^2$ CNR-IOM Democritos National Simulation Center, Via Bonomea 265, I-34136 Trieste, Italy\\
$^3$ International Centre for Theoretical Physics (ICTP), P.O. Box 586, I-34014 Trieste, Italy\\
$^4$ NEST, Scuola Normale Superiore and Istituto Nanoscienze-CNR, I-56126 Pisa, Italy 
}

\begin{abstract}
We revisit here the issue of thermally assisted Quantum Annealing by a detailed study of the dissipative Landau-Zener problem
in presence of a Caldeira-Leggett bath of harmonic oscillators, 
using both a weak-coupling quantum master equation and a quasi-adiabatic path-integral approach.    
Building on the known zero-temperature exact results (Wubs {\em et al.}, PRL {\bf 97}, 200404 (2006)), we show that 
a finite temperature bath can have a beneficial effect on the ground-state probability only if it couples also to a spin-direction that 
is {\em transverse} with respect to the driving field, while no improvement is obtained for the more commonly studied 
purely longitudinal coupling. 
In particular, we also highlight that, for a transverse coupling, raising the bath temperature further improves the ground-state 
probability in the fast-driving regime. 
We discuss the relevance of these findings for the current quantum-annealing flux qubit chips.
\end{abstract}
\maketitle

\section{Introduction} \label{sec:intro}
%

Quantum Annealing (QA) \cite{Finnila_CPL94,Kadowaki_PRE98,Brooke_SCI99,Santoro_SCI02}
--- essentially equivalent to a form of quantum computation known as Adiabatic Quantum Computation (AQC) \cite{Farhi_SCI01} ---
was originally introduced as an alternative to classical simulated annealing \cite{Kirkpatrick_SCI83}
for optimization and has  become a very active field of research in the last few years,
due to the availability of QA programmable machines based on superconducting flux quantum bits \cite{Harris_PRB10,Johnson_Nat11}.

One of the open issues in this field is the role played by a thermal environment, which makes the dynamics of the system 
non-unitary \cite{Ashhab_PRA06,Patane_PRL08,Max_arXiv17,Malla_arXiv17}. 
It has been argued \cite{Amin_PRA09} that AQC-QA should be less critically affected by decoherence/dephasing 
of the individual qubits with respect to traditional unitary-gate quantum computation.
Even more, it has been suggested \cite{Amin_PRL08} that  a thermal environment might be
{\em beneficial} during the open system dynamics, in the sense that it might enhance the probability that the system 
remains in its instantaneous ground state with respect to the corresponding coherent evolution dynamics. 
This mechanism is known as ``thermally assisted AQC''  \cite{Amin_PRL08}.

A way to model the coupling of the quantum system with the external environment is to write a system-plus-bath Hamiltonian of the form 
\cite{Leggett_RMP87,Weiss:book,Breuer:book,Grifoni_PR98}:
\begin{equation}
\hat{H}(t) = \hat{H}_{\rm sys}(t) + \sum_{\nu} \hat{A}_{\nu} \otimes \hat{X}_{\nu} +
\hat{H}_{\rm B} \;,
\end{equation}
where $\hat{H}_{\rm sys}(t)$ is the Hamiltonian of the time-dependent quantum system, 
$\hat{A}_{\nu}$ and $\hat{X}_{\nu}$ denote a suitable set of Hermitean operators acting on the system and on the bath respectively,
and $\hat{H}_{\rm B}$ describes a set of harmonic oscillators.
For a single qubit (spin-1/2), one would take $\hat{A}_{\nu} = \PauliSigma^{\nu}$ (the Pauli matrices) and 
$\hat{X}_{\nu}$ combinations of the bosonic harmonic oscillator operators.   
As a matter of fact, both in the spin-boson literature as well as in the context of QA, a $\PauliSigma^z$-coupled
environment is usually assumed.
Indeed, if $\hat H_{\rm sys}$ describes a biased spin-boson problem \cite{Leggett_RMP87},
$\hat{H}_{\rm sys} = -(\epsilon/2) \PauliSigma^z - (\Delta/2) \PauliSigma^x$, 
then $\PauliSigma^z \otimes \hat{X}_z$ is a rather ``generic'' linear coupling to the environment,
as it causes both decoherence and relaxation of the density matrix populations \cite{Schoen_PhyScr02}. 
Moreover, as discussed in Ref.~\onlinecite{Leggett_RMP87}, the $\PauliSigma^z$-coupling is definitely the 
dominant noise mechanism when the qubit originates from a {\em macroscopic} two-level system, as 
any off-diagonal coupling would be proportional to the ``exponentially small'' (overlap-related) 
tunnelling matrix element $\Delta$.
In the context of QA, the $\PauliSigma^z$-noise is recognized to be the dominant noise mechanism for the 
D-Wave$^{\mbox{\tiny \textregistered}}$ machine superconducting flux qubits \cite{Harris_PRB10,Johnson_Nat11}, 
albeit with important low-frequency noise contributions which tend to make the qubits dynamics typically 
dominated by incoherent tunnelling \cite{Harris_PRL08,Amin_MRT_PRL08}.
  
While a general AQC-QA Hamiltonian $\hat{H}_{\rm sys}(t)$ would involve a complex dynamics of exponentially many states, 
a simple dissipative two-level system serves as a pedagogical example of the effect of the environment on the 
avoided-level crossing of the two lowest-lying adiabatic states of $\hat{H}_{\rm sys}(t)$.
Even more, a 16-qubit problem has been specially engineered on the D-Wave$^{\mbox{\tiny \textregistered}}$ chip \cite{Dickson_NatCom13} 
in such a way that the two lowest instantaneous eigenstates, which are quite well separated from the higher states,
provide an explicit realization of a dissipative Landau-Zener (LZ) problem, for which one would write the system Hamiltonian 
$\hat{H}_{\rm sys}(t) \equiv \hat{H}_{\rm Q}(t)$ as
\begin{equation} 	\label{eqn:LZ_model}
	\hat{H}_{\rm Q}(t) = -\frac{\epsilon(t)}{2} \PauliSigma^z - \frac{\Delta}{2}\PauliSigma^x \;,
\end{equation}
with $\epsilon(t) = vt$ the driving field, and $v$ the driving velocity.
This Hamiltonian describes the avoided crossing between the two eigenstates of $\PauliSigma^z$,
$\{|\!\!\downarrow \rangle, |\!\!\uparrow \rangle\}$, for  $t \to \pm\infty$, due to the coupling provided by an off-diagonal term 
$\sim \Delta \, \PauliSigma^x$.
From general considerations, when the LZ physics emerges from the two lowest-lying eigenstates 
$\{|\psi_1(t)\rangle, |\psi_2(t)\rangle\}$ of a complex multi-qubit Hamiltonian, one would expect that an appropriate
model to describe the dissipation should include couplings to the transverse directions, {\it e.g.}, 
\begin{equation} \label{eqn:dissLZ_full}
\hat{H}(t) = \hat{H}_{\rm Q}(t) + \left( \sum_{\nu} g_{\nu} \, \PauliSigma^{\nu} \right) \otimes \hat{X} + \hat{H}_{\rm B} \;.
\end{equation}
Here the sum runs over $\nu=x,y,z$ so as to allow for the possibility of a coupling (in principle time-dependent)
through $g_{x/y} \PauliSigma^{x/y}$, which we will refer to as {\em transversal} noise, as well as {\it via} the more standard {\em longitudinal} coupling $g_z\PauliSigma^z$. 
Notice that, although in principle one might allow for independent harmonic oscillator baths coupling to each spin direction $\PauliSigma^{\nu}$, for the purpose of a simpler setting we focus here on a common set of oscillators, \textit{i.e.} $\hat{X}_\nu = \hat{X}$.
Couplings of this kind have already been considered in the literature, precisely in the present dissipative LZ context
\cite{Wubs_06,Saito_07,Nalbach_09,Nalbach_PRA14,Javanbakht_15}.

A number of results are known on this problem, both exact \cite{Wubs_06, Saito_07} at zero temperature and numerical 
at finite temperature \cite{Nalbach_09,Nalbach_PRA14,Javanbakht_15}, 
obtained by means of a perturbative {\em quantum master equation} (QME) approach and by the so-called {\em quasi-adiabatic path integral} 
(QUAPI) \cite{Makri_95,Makri_95_bis}. 
Let $P_{\gs}(v,T)$ denote the probability that the system remains in the ground state of $\hat{H}_{\rm Q}(t)$ when evolving
from the ground state at $t=-\infty$ up to $t=+\infty$ in presence of a thermal bath at temperature $T$. 
It is known that a $\PauliSigma^z$-coupling alone cannot be truly beneficial to $P_{\gs}(v,T)$. 
This has been established both exactly, at zero temperature 
--- where the actual $P_{\gs}{(v,T=0)}$ is completely unaffected by the bath \cite{Saito_07} and coincides with the
well-known Landau-Zener \cite{Landau_PZS32_2,Zener_PRS32} coherent evolution result 
$P^{\LZ}_{\gs}(v)=1-\nep^{-\pi\Delta^2/(2\hbar v)}$ ---,  
and numerically at finite temperature\cite{Nalbach_09,Javanbakht_15}. 

\begin{figure}
\begin{center}
  \includegraphics[width=\columnwidth]{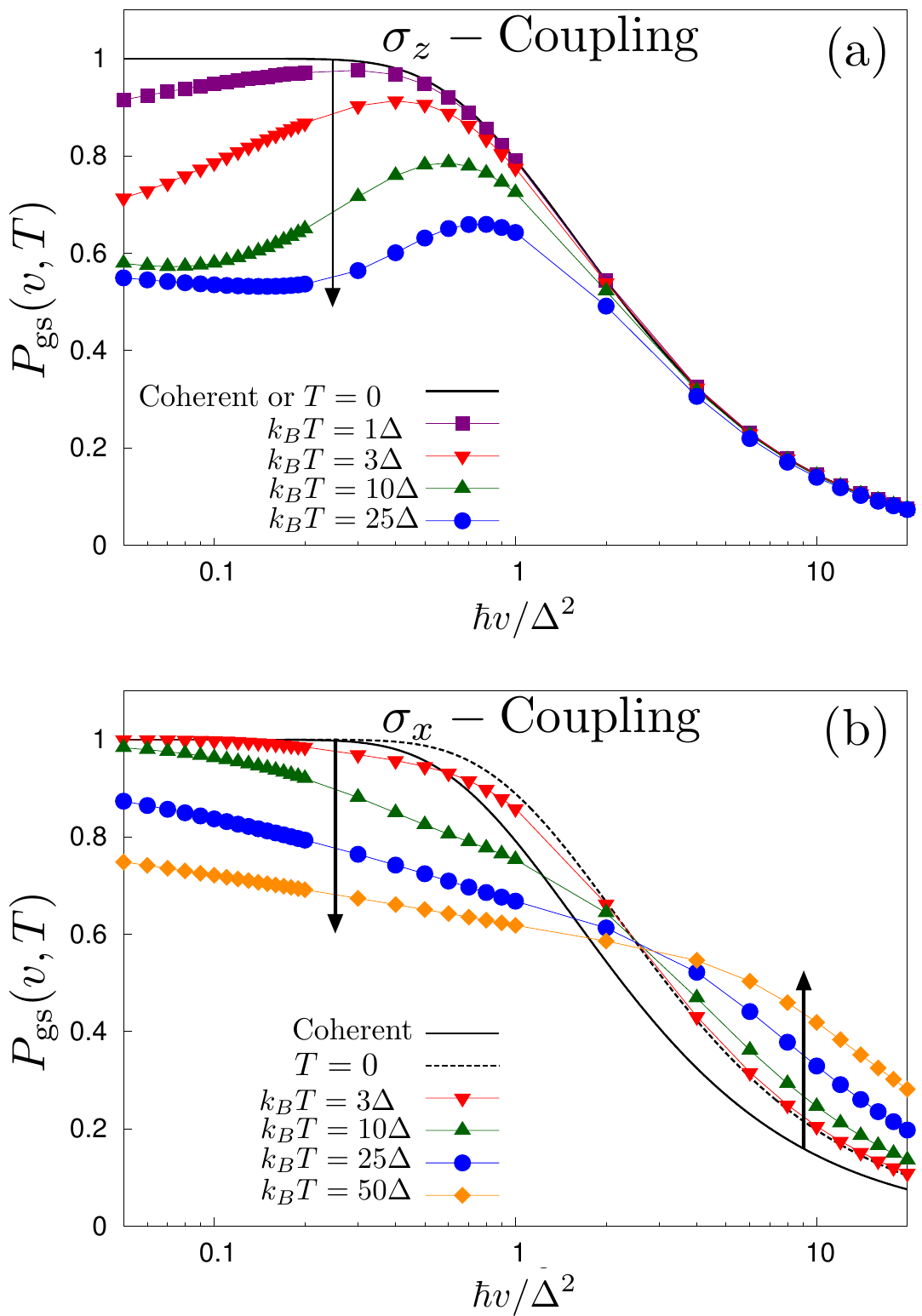} 
\end{center}
\caption{The probability $P_{\gs}(v,T)$ of remaining in the ground state of the Hamiltonian $\hat{H}_{\rm Q}(t)$ when evolving with
Eq.~\eqref{eqn:dissLZ_full} in presence of a  thermal bath at temperature $T$ {\it versus} the driving velocity $\hbar v/\Delta^2$, 
for a purely longitudinal $\PauliSigma^z$-coupling (a) and for a purely transversal $\PauliSigma^x$-coupling (b) to the environment. 
The bath properties are fully determined by the choice of its spectral function; here we have considered an ohmic bath with a cutoff frequency 
$\omega_c=10\Delta/\hbar$ and a coupling $\alpha=2 \times 10^{-3}$  (see Sec.~\ref{sec:model} for precise definitions).  
The black solid line is the coherent-evolution result $P_{\gs}^{\LZ}(v)$, the black dashed line in  panel (b) is the probability 
$P_{\gs}(v,T=0)$ obtained by means of the exact analytical result at zero temperature\cite{Saito_07}, 
while the various points are QME data for different bath temperatures. 
Black arrows indicate the direction of increasing $T$.
}
\label{fig:summary}
\end{figure}

In presence of a transverse coupling, the situation changes drastically:
an exact analysis \cite{Wubs_06,Saito_07} at $T=0$ shows that $P_{\gs}(v,T=0)$ can be enhanced with respect to 
the coherent probability $P_{\gs}^{\LZ}(v)$. 
On the contrary, the finite-$T$ behaviour of $P_{\gs}(v,T)$, and the possibility of a ``thermally assisted'' QA, {\it i.e.}, 
a beneficial effect due to the bath, has not been properly scrutinized. 
This is what we aim at exploring in the present paper.
We will show that the behaviour of $P_{\gs}(v,T)$ in presence of a transverse coupling becomes highly non-trivial upon rising $T$.
In particular, when the coupling to the bath is along $\PauliSigma^x$, while increasing $T$ 
{\em reduces} $P_{\gs}(v,T)$ in the adiabatic regime of small $\hbar v/\Delta^2$, it further {\em improves} on $P_{\gs}(v,T=0)$ 
for fast drivings, \textit{i.e.} in the non-adiabatic regime of large $\hbar v/\Delta^2$. 
Part of the story and some of our findings are summarized in Fig.~\ref{fig:summary}, 
where $P_{\gs}(v,T)$ is plotted as a function of $\hbar v/\Delta^2$ for both a 
$\PauliSigma^z$- and  a $\PauliSigma^x$-coupling. 
These results will allow us to partly address the physics behind the experimental findings of Ref.~\onlinecite{Dickson_NatCom13}, 
which we will comment further upon in the final discussion. 

The paper is organized as follows.
In Sec.~\ref{sec:model} we discuss the properties of the bosonic bath to which the qubit is coupled and we quickly survey the techniques used:  
the QME approach and the QUAPI technique. 
In Sec.~\ref{sec:results} we present our results. 
We have benchmarked the validity of our QME approach both at zero temperature, by comparing it with known exact analytical results as well as at finite $T$, against numerically exact QUAPI results.
In passing, we find a remarkable and perplexing robustness of our QME for a $\PauliSigma^x$-coupled bath at $T=0$ up to the strong-coupling 
and fast driving regimes.
The QUAPI technique has also been used to address the strong-coupling regime for a longitudinal coupling.  
We also discuss a simple single-oscillator-bath model in terms of which the effect of finite temperature for a $\PauliSigma^x$-coupling 
becomes physically very transparent. 
Finally, Sec.~\ref{sec:conclusions} contains a discussion and some concluding remarks. 

\section{Model and methods} \label{sec:model}

The model considered in this work, see Eqs. (\ref{eqn:LZ_model}-\ref{eqn:dissLZ_full}), 
has been extensively discussed in the literature \cite{Leggett_RMP87,Gefen_PRB87,AoRammer_PRB91,Weiss:book,Breuer:book,Grifoni_PR98}. 
For this reason, we just mention a few details, for the reader's convenience, concerning the properties of the 
bath and the choice of the initial state.
Specifically, the Hamiltonian we will study is given by:
\begin{equation} \label{eqn:dissLZ_xz}
\hat{H}(t) = \hat{H}_{\rm Q}(t) + \frac{1}{2}\left( \cos{\theta} \, \PauliSigma^z + \sin{\theta} \, \PauliSigma^x \right) \otimes \hat{X} 
+ \hat{H}_{\rm B} \;,
\end{equation}
where we set for simplicity $g_y=0$ and parameterize,
following Ref.~\onlinecite{Saito_07}, the transversal $g_x$ and longitudinal $g_z$ couplings as
$(g_x, g_z) = (\sin\theta, \cos\theta )/2$ (neglecting also any time-dependence of the couplings).
The bath Hamiltonian $\hat{H}_{\rm B}=\sum_{k} \hbar\omega_{k} \opbdag{k}\opb{k}$ describes a set of harmonic oscillators, 
$[\opb{k},\opbdag{k'}]=\delta_{k,k'}$,
and $\hat{X}=\sum_{k} \lambda_{k} (\opbdag{k}+\opb{k})$ is a combination of their position operators,
with coupling constants $\lambda_k$.
For a bath of harmonic oscillators, the coupling to the environment is captured by the spectral function  
$J(\omega) = \sum_{k} \lambda_{k}^2 \delta (\omega-\omega_{k})$ which we take to be parametrized, in
the limit of a continuous distribution of frequencies $\omega_{k}$, by the standard ohmic form \cite{Leggett_RMP87}
$J(\omega) = 2\alpha \hbar^2\omega \nep^{-\omega/\omega_c}$, with a cut-off frequency $\omega_c$ and
an overall dimensionless coupling constant $\alpha$, which encodes the coupling strength between the system and the environment.
Since our results are not qualitatively affected by the choice of the cut-off frequency, 
we will set $\omega_c=10 \Delta/\hbar$ throughout this paper.

Let us briefly discuss how the initial state of the system is chosen. 
In the spirit of the LZ formula, we assume that, at some large negative time $t_0=-t_a$ 
--- which we select such that $v t_a\gg \Delta$, so that the initial qubit ground state $|\psi_{\gs}(t_0)\rangle$ is very close to being a 
$\PauliSigma^z$-eigenstate ---
we initialize the system in a decoupled state $\hat{\rho}_0 = |\psi_{\gs}(t_0)\rangle \langle \psi_{\gs}(t_0)| \otimes \hat{\rho}_{\rm B}$,
where $\hat{\rho}_{\rm B} = \nep^{-\beta \hat{H}_{\rm B}}/Z$ is the bath thermal equilibrium density matrix
at temperature $T$, with $Z=\Tr (\nep^{-\beta \hat{H}_{\rm B}})$ the bath partition function and $\beta=1/(k_{ B}T)$.
The ensuing unitary dynamics is captured by the full evolution operator
\begin{equation} \label{time_evolut}
	\hat{\mathcal{U}}(t,t_0)= \Texp \left( - \frac{i}{\hbar}\int_{t_0}^t dt' \; \hat{H}(t') \right) 
\end{equation}
where $\Texp$ is the time-ordered exponential. 
Information on the qubit is fully encoded in its reduced density matrix
\begin{equation} \label{qubit_density_matrix_evolved}
	\hat{\rho}_{\rm Q}(t)= \Tr_{\rm B} \left( \hat{\mathcal{U}}(t,t_0) \hat{\rho}_0 \, \hat{\mathcal{U}}^\dagger (t,t_0) \right) 
\end{equation}
and we can extract the probability that the qubit can be found in its (instantaneous) ground state $|\psi_{\gs}(t)\rangle$ at time $t$:
\begin{equation} \label{eqn:Pgs_t}
{\mathrm P}_{\gs}(t)=\Tr_{\rm Q} \Big( \hat{\rho}_{\rm Q}(t)\; |\psi_{\gs}(t)\rangle \langle \psi_{\gs}(t) |  \Big) \;.
\end{equation}
From the probability ${\mathrm P}_{\gs}(t)$ calculated at a large positive time $t_{\rm f}=-t_0=t_a$, we obtain the quantity that effectively
generalizes the LZ formula, \textit{i.e.} $P_{\gs}(v,T)\equiv {\mathrm P}_{\gs}(t_{\rm f})$, which turns out to be effectively 
independent of the value of $t_a$, provided $vt_a \gg \Delta$.
In practice, choosing $vt_a =200\Delta$ is enough to guarantee convergence to the infinite-time limit, 
and we have set it throughout this paper, unless otherwise stated.
In the following, we will compare results obtained with $\hat{\rho}_{\rm Q}(t)$ calculated in two different ways: 
first, by means of a rather standard QME approach based on Bloch-Redfield-type equations, see Sec.~\ref{QME}; 
second, by using a non-perturbative, non-Markovian numerical method, the quasi-adiabatic path-integral (QUAPI), 
see Sec.~\ref{QUAPI}, which we employed to check and benchmark the QME data and to produce reliable 
results in the strong-coupling limit. 

\subsection{Quantum Master Equation} \label{QME}
We construct our QME for a system that is weakly coupled to the thermal environment, 
assuming the usual Born-Markov approximations and proceeding as in many standard derivations \cite{Cohen:book,Gaspard_JCP99a,Yamaguchi_PRE17}. 
We provide a few details about this derivation in Appendix \ref{appA}.
In order to make a more direct contact with the well known weak-coupling QME for an unbiased 
spin-boson problem ($\epsilon=0$), we choose to perform, following Ref.~\onlinecite{Nalbach_PRA14}, 
a time-dependent rotation in spin space around the $y$-axis, $\hat{R}_t = \exp[i\phi_t\PauliSigma^y/2]$,
with $\phi_t = \arctan(\epsilon(t)/\Delta)$, such that
$\hat{R}_t^\dagger \hat{H}_{\rm Q}(t) \hat{R}_t = -E_t \PauliSigma^x / 2$, 
with $E_t \equiv \hbar\Lambda_t \equiv \sqrt{\Delta^2 +\epsilon^2(t)}$.
In this framework, by adopting the Bloch-sphere representation 
$\tilde{\rho}_{\rm Q}(t) = \hat{R}_t^\dagger \hat{\rho}_{\rm Q}(t) \hat{R}_t = \frac{1}{2} (\id + \sum_{\nu}r_{\nu}(t)  \PauliSigma^{\nu})$ with $\nu=x,y,z$,
we write down our weak-coupling QME as the following set of three differential equations:
\begin{equation} \label{QME_noRWA}
\left\{ \begin{array}{lcl}
	\dot{r}_x &=& -\gamma_{\Relax} (r_x - \overline{r}_x) + (\dot{\phi}_t + \gamma_{xz}) r_z \\
	\dot{r}_y &=& -\left(\gamma_{\Deco} + \displaystyle \frac{\gamma_{\Relax}}{2}\right )\, r_y + \Lambda_t r_z \\
	\dot{r}_z &=& -\dot{\phi}_t r_x -\gamma_{zx}(r_x - \overline{r}_x) - \Lambda_t r_y 
                           -\displaystyle \left(\gamma_{\Deco} - \frac{\gamma_{\Relax}}{2} \right) \, r_z 
\end{array}
\right.
\end{equation}
with $\overline{r}_x(t) = \tanh(\beta E_t / 2)$  the ``instantaneous'' putative equilibrium value that $r_x$ would 
reach in absence of driving. 
The various (time-dependent) rate constants include the usual 
``relaxation'' $\gamma_{\Relax}$, ``pure dephasing'' $\gamma_{\Deph}$ and 
``decoherence'' $\gamma_{\Deco}$ rates \cite{Schoen_PhyScr02}:
\begin{subequations}
\begin{align}
	&\gamma_{\Relax}(t) = \frac{\pi}{2 \hbar^2} \coth \left(\frac{\beta \hbar \Lambda_t}{2}\right) J(\Lambda_t) 
\, \cos^2(\phi_t+\theta) 
\\
	&\gamma_{\Deph}(t) = \frac{2\pi\alpha}{\hbar\beta} \, \sin^2(\phi_t+\theta)
\\
&\gamma_{\Deco}(t) = \gamma_{\Deph}(t)+\frac{1}{2} \gamma_{\Relax}(t)  \; ;
\end{align}
\end{subequations}
as well as the following two extra terms:
\begin{subequations}
\begin{align}
&\gamma_{zx}(t) = -\frac{\pi}{4 \hbar^2} \coth \left(\frac{\beta \hbar \Lambda_t}{2}\right) J(\Lambda_t) \, \sin 2(\phi_t+\theta) \\	
&\gamma_{xz}(t) = \frac{\pi\alpha}{\hbar\beta} \, \sin 2(\phi_t+\theta) \;.
\end{align}
\end{subequations}
These equations are obtained without using the so-called {\em secular} or {\em rotating wave approximation} (RWA) \cite{Cohen:book}; 
if we employ it, we obtain a simplified set of equations \cite{Nalbach_PRA14,Javanbakht_15} which we report in Appendix \ref{appB},
see Eq.~\eqref{QME_RWA}. 
In the following, we will present results obtained from Eqs.~\eqref{QME_noRWA}, integrated through a standard  IV-order Runge-Kutta method.
We will comment, in Appendix \ref{appB}, on the quality of the RWA-approximation in some regimes. 

\subsection{QUAPI} \label{QUAPI}
The quasi-adiabatic propagator path-integral algorithm (QUAPI) developed by Makri and 
Makarov \cite{Makri_95,Makri_95_bis,Makri_95_tris} is a numerical technique for evaluating an exact Trotter-discretized 
path-integral of a low-dimensional system (the evolving qubit $\hat{H}_{\rm Q}(t)$, in the present case)
taking into account the effect of the bath through a discrete version of the Feynman-Vernon influence integral, describing
non-local-in-time correlations induced by the environment. 
The key idea behind this method is that the resulting path-integral is evaluated without resorting to any Monte Carlo sampling,
but rather through an ingenious use of an iterative tensor multiplication scheme. 
This technique is numerically exact, in principle, since both the Trotter time-step $\delta t$ and the size of the tensor
$2^K$ --- where $K \geq 1$ is such that $\tau_{\rm mem} = K \delta t$ sets the bath-memory time-scale, 
beyond which the bath correlation function is effectively neglected --- can be in principle
changed towards their limiting values of $\delta t\to 0$ and $K\to \infty$ for each fixed $\tau_{\rm mem}$. 
In practice, however, both $\delta t$ and $K$ have to be chosen carefully. 
For large $\delta t$, the required $K$ is small, and one deals with small matrices, but the Trotter error might be large; 
for small $\delta t$ the Trotter error might be negligible, but one needs large values of $K$, hence the resulting matrices 
are big and the calculation becomes very heavy.

\section{Results} \label{sec:results}
%
\subsection{Zero temperature}
Let us start by discussing our results at zero-temperature. In this regime, we have a perfect benchmark for our QME approach 
given by the exact predictions by H\"anggi and coworkers \cite{Wubs_06,Saito_07}, who showed that, in presence of a bath 
at $T=0$ and for an evolution starting at $t_0=-\infty$ and ending at $t_{\rm f}=+\infty$, the exact generalization of the LZ formula reads:
\begin{equation} \label{T0_Saito_formula}
	P_{\gs}(v,T=0) = 1 - \nep^{-\frac{\pi W_{\theta}^2}{2\hbar v}}
\end{equation}
where $W_{\theta}$, effectively replacing the tunnelling matrix element $\Delta$ in the standard LZ formula, 
for our ohmic bath choice reads:
\begin{equation} \label{eqn:ohmic_W}
	W_{\theta}^2 = \Big| \Delta - \alpha \hbar\omega_c \sin{2\theta} \Big|^2 + 2\alpha(\hbar\omega_c)^2 \sin^2\theta \;.
\end{equation}
%
\begin{figure}
\begin{center}
  \includegraphics[width=\columnwidth]{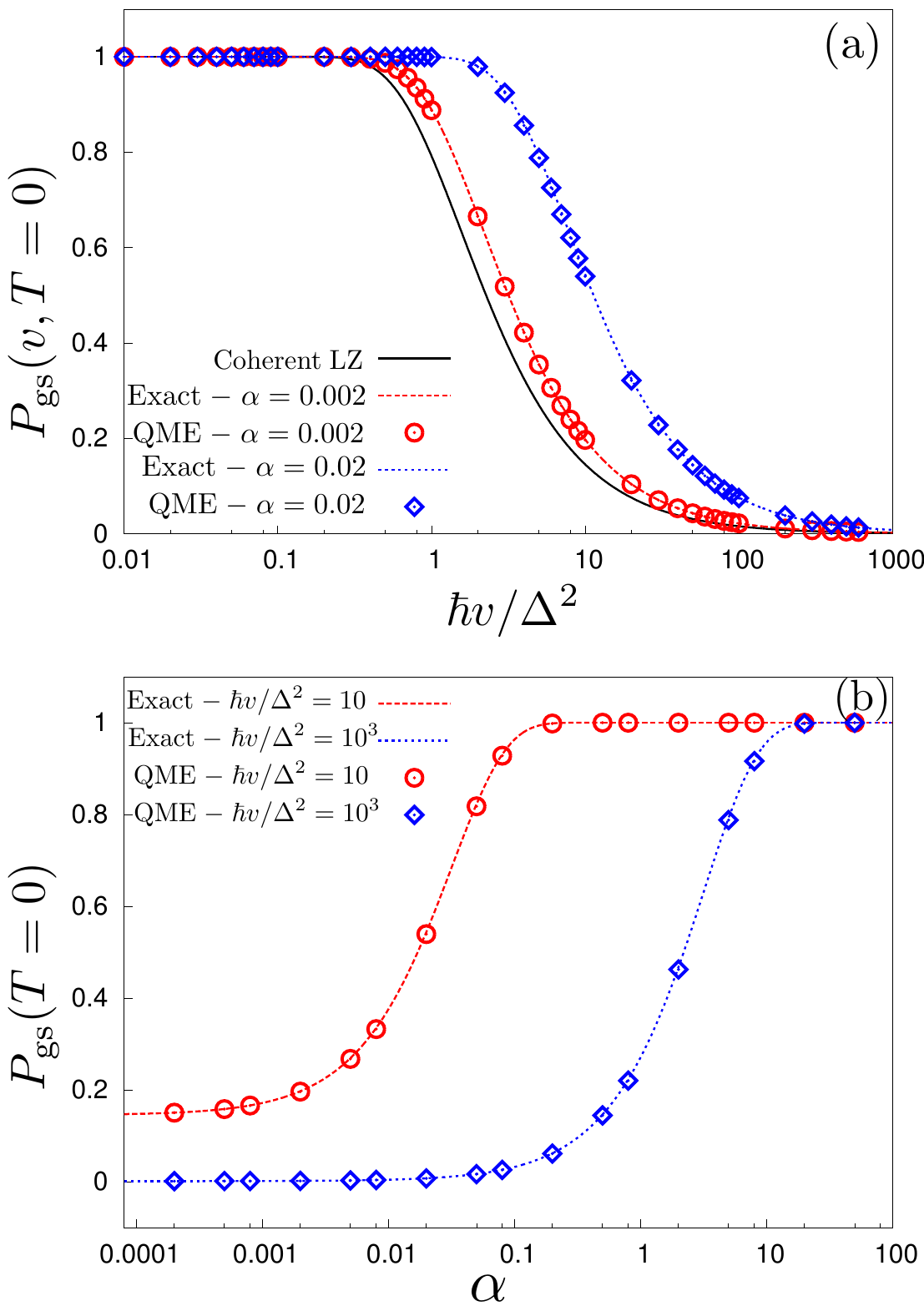} 
\end{center}
\caption{Check of the validity of the QME approach at zero temperature for a pure transversal noise, $\theta=\pi/2$. 
The probability to follow the ground state at zero temperature $P_{\gs}(T=0)$ {\it versus} (a) the driving velocity $\hbar v/\Delta^2$ 
and (b) the coupling strength $\alpha$.
The lines are the exact predictions obtained using Eq.~\eqref{T0_Saito_formula}, while the points correspond to our QME in  Eqs.~\eqref{QME_noRWA} (without RWA). Here $\hbar\omega_c=10\Delta$.}
\label{fig:PT0_sigmax}
\end{figure}

\begin{figure}
\begin{center}
\includegraphics[width=\columnwidth]{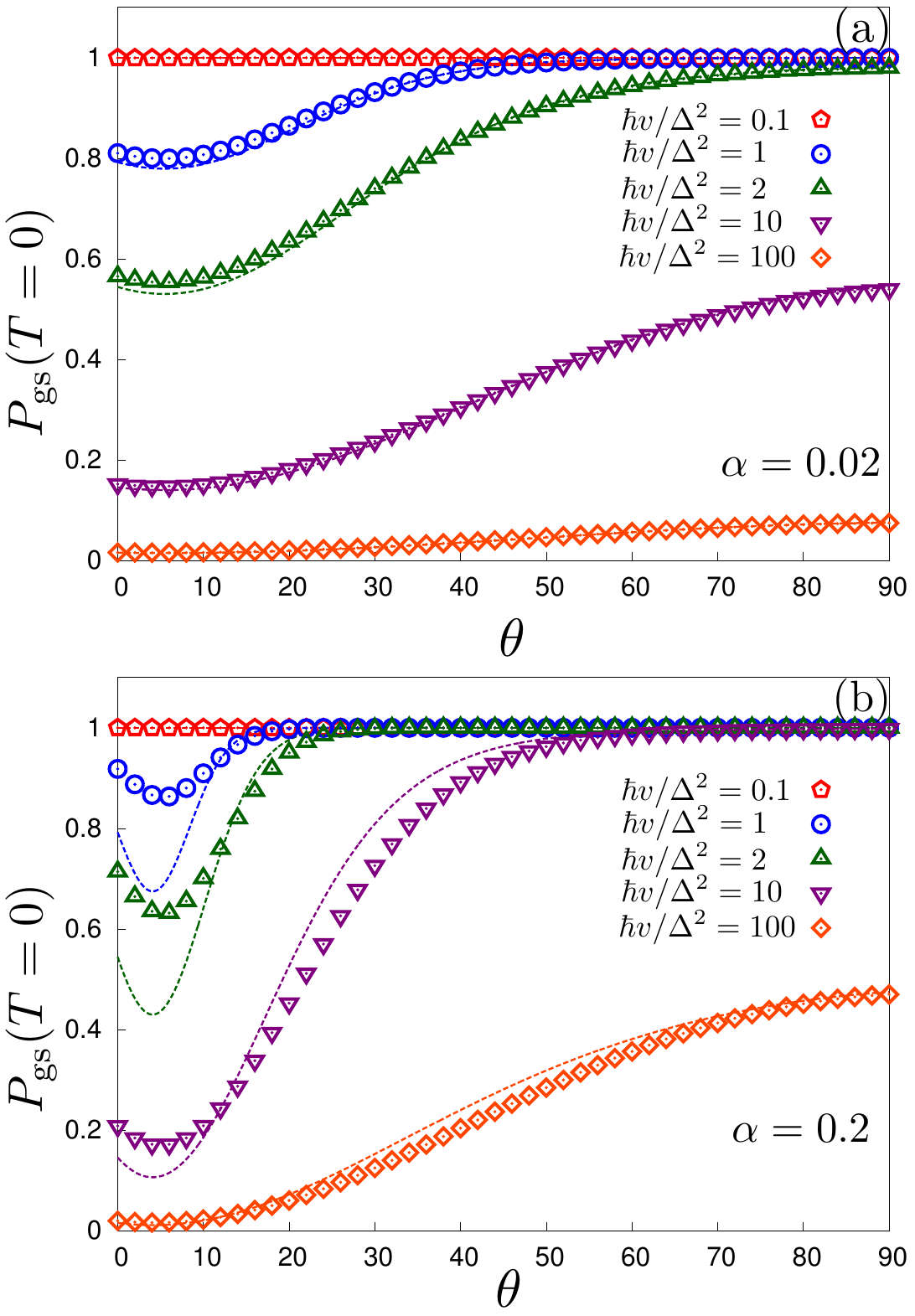}
\end{center}
\caption{Check of the validity of the QME approach at zero temperature.
Probability to follow the ground state $P_{\gs}(v,T=0)$ for fixed driving velocities $\hbar v/\Delta^2$ {\it versus} the noise coupling 
direction $\theta$ for $\alpha=0.02$ (a), and  $\alpha=0.2$ (b).
Lines are the exact results of Ref.~\onlinecite{Saito_07}, points correspond to our QME in Eqs.~\eqref{QME_noRWA}.
}
\label{fig:PT0_theta}
\end{figure}

According to Eq. \eqref{eqn:ohmic_W},  the bath has  no effect when acting only along $\PauliSigma^z$
since $W_{\theta=0}^2=\Delta^2$; on the contrary, it effectively increases the bare tunnelling amplitude $\Delta$ when, for
instance, it acts along $\PauliSigma^x$ since $W_{\theta = \pi/2}^2 = \Delta^2 + 2\alpha\omega_c^2$.
(Incidentally, an identical enhancement would hold for a bath coupling along $\PauliSigma^y$.)
This effective increase of the tunnelling amplitude leads to a definite {\em enhancement} of the probability to remain
in the ground state $P_{\gs}(v,T=0)$: we illustrate this in Fig.~\ref{fig:PT0_sigmax}(a), where the exact predictions
of Refs.~\onlinecite{Wubs_06,Saito_07} reported in Eq.~\eqref{T0_Saito_formula} are compared to QME evolution data,
obtained by integrating Eqs.~\eqref{QME_noRWA} for two relatively weak values of the coupling $\alpha$.
The agreement is almost perfect at weak coupling, and, as shown in Fig.~\ref{fig:PT0_sigmax}(b),
it also persists in a remarkable and puzzling fashion all the way up to the strong coupling regime. 
Therefore, our QME is extremely good for a pure transversal noise,
$\theta=\pi/2$, even in the strong coupling regime. 
This excellent agreement diminishes, or even disappears, if the noise has a longitudinal component, as shown in Fig.~\ref{fig:PT0_theta}, 
where the probability $P_{\gs}(v,T=0)$ is plotted, for fixed driving velocities $v$, {\it versus} the noise coupling direction $\theta$ 
and for different coupling strengths $\alpha$. 
The reason for this great reliability is still unclear to us, but we speculate that it might be linked to the fact that the system's dynamics is Markovian 
for a purely transversal coupling, while non-Markovian effects are more pronounced for a purely longitudinal noise \cite{Breuer_RMP16}.
Indeed, we have checked the behaviour in time of the trace distance between pairs of random initial states both for $\theta = 0$ and $\theta = \pi/2$: 
in the former case, we have found non-monotonic trends, hinting that the dynamics must be non-Markovian \cite{Breuer_RMP16}; 
in the latter case, we have found that the trace distance always monotonically decreases, as expected for Markovian dynamics \cite{Breuer_RMP16}.
%

\subsection{Finite temperature}
We now turn to finite bath temperatures. 
We begin by considering a purely longitudinal coupling, $\theta=0$. 
At $T>0$ no analytical results are available, but we can use QUAPI to benchmark our QME data.
%
\begin{figure*}
\begin{center}
  \includegraphics[width=175mm]{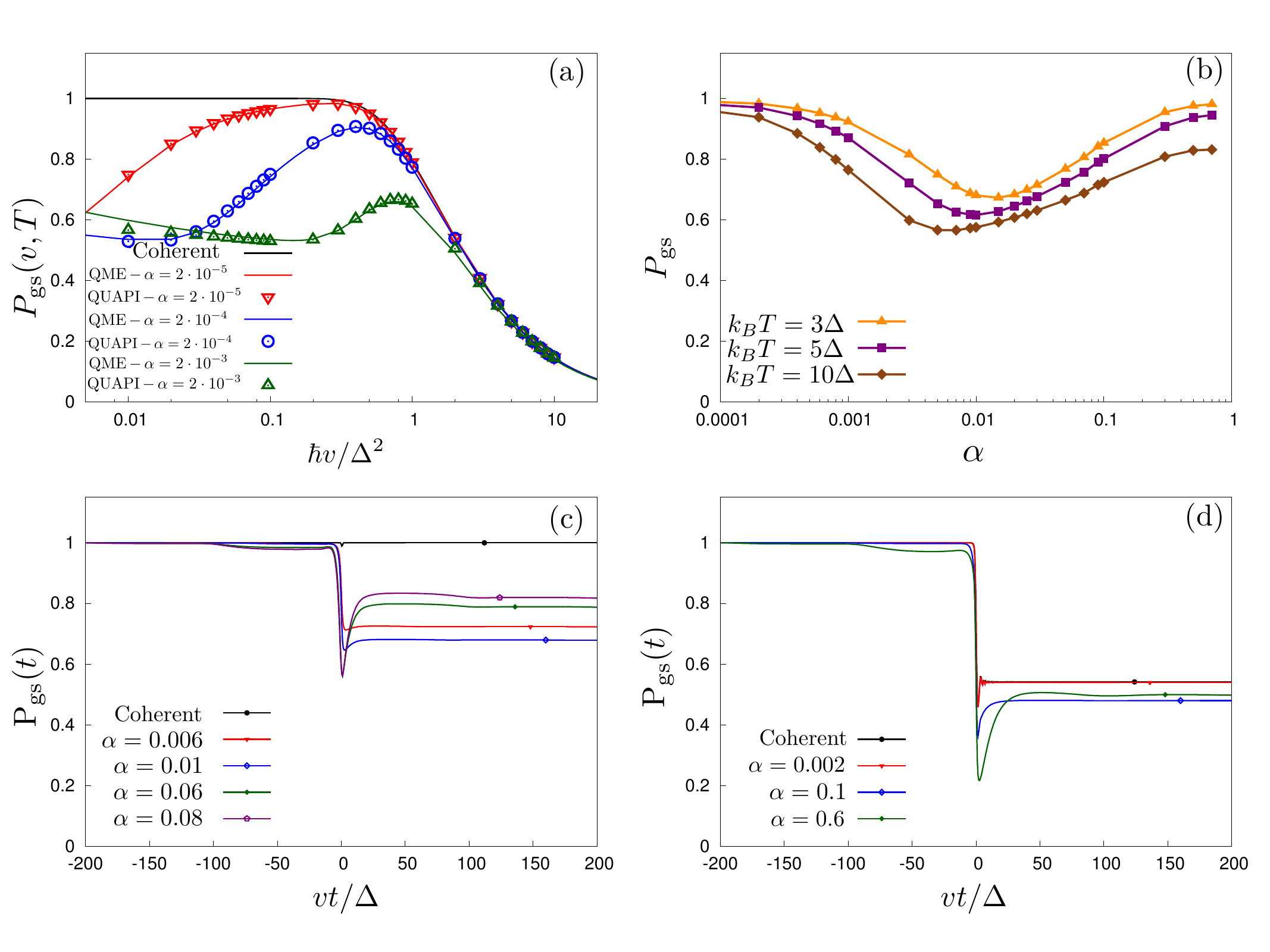}
\end{center}
\caption{(a) $P_{\gs}(v,T)$ as a function of the driving $\hbar v/\Delta^2$ for pure longitudinal noise, $\theta=0$, 
at finite temperature $k_BT=25\Delta$ and different values of the ohmic coupling strength $\alpha$. 
Lines correspond to QME results, points to QUAPI. 
The black solid line shows, as a guide to the eye, $P_{\gs}^{\LZ}(v)\equiv P_{\gs}(v,T=0)$.
(b) $P_{\gs}$ {\it versus} the ohmic coupling strength at a fixed driving velocity $\hbar v=0.2\Delta^2$ and different bath temperatures.
(c,d) Detail of the time-evolving ${\mathrm P}_{\gs}(t)$ for ``slow'' (c, $\hbar v/\Delta^2=0.2$) and ``fast'' (d, $\hbar v/\Delta^2=2$)
drivings, for various couplings $\alpha$ and $k_BT=3\Delta$.
Data in (b-d) obtained using QUAPI, and $\hbar\omega_c=10\Delta$ throughout. 
}
\label{fig:PvT25}
\end{figure*}
%
As Fig.~\ref{fig:PvT25}(a) shows, the agreement between the QME results (lines) and the QUAPI data (points)
is quite good in the weak coupling regime and (not shown) for all the temperatures for which a QUAPI simulation is feasible.
Notice also that increasing the bath temperature reduces $P_{\gs}(v,T)$ from the coherent evolution and $T=0$ value 
$P_{\gs}(v,0)=P_{\gs}^{\LZ}(v)$, and it does so in a rather non-monotonic fashion, 
depending on the coupling constant $\alpha$, as previously reported \cite{Nalbach_09,Nalbach_PRA14}. 
What we also find is that $P_{\gs}(v,T)$ exhibits a non-monotonic behavior for increasing coupling $\alpha$, at fixed
$v$, especially relevant in the adiabatic driving regime (small $\hbar v/\Delta^2$). 
This is shown in Fig.~\ref{fig:PvT25}(b), where we plot $P_{\gs}(v,T)$ at a fixed value of $\hbar v/\Delta^2=0.2$ and 
for different temperatures, as a function of the coupling constant $\alpha$ (these data are obtained using QUAPI, as
the QME would not be reliable at such large values of $\alpha$ for $\theta=0$). 
While for weak coupling the probability $P_{\gs}(v,T)$ decreases as $\alpha$ increases, the inverse tendency, characterized by
a $P_{\gs}(v,T)$ increasing back towards $P_{\gs}^{\LZ}$, is found when $\alpha$ increases beyond a certain 
$T$-dependent characteristic $\alpha_{\rm min}(T)$.
Evidently, although the dynamics tends to become more and more incoherent for large $\alpha$, the overall form of 
$P_{\gs}(v,T)$ would be very close to the fully coherent value of $P_{\gs}^{\LZ}$, in agreement with the Fermi-Golden rule results of Ref.~\onlinecite{Amin_MRT_PRL08}, obtained from a fully incoherent population-dynamics evolution.
Figure~\ref{fig:PvT25}(c,d) shows details of the dynamical evolution of ${\mathrm P}_{\gs}(t)$ obtained from QUAPI at
different values of the parameters. In Fig.~\ref{fig:PvT25}(c) the driving velocity is ``small'' ($\hbar v/\Delta^2=0.2$), the coherent evolution
is essentially adiabatic, and we notice the effect of the finite-temperature bath in the form of a rather sharp decrease
of ${\mathrm P}_{\gs}(t)$ around the transition region, followed by a partial recovery which tends to push ${\mathrm P}_{\gs}(t_{\rm f})$
towards values which are closer and closer to $P_{\gs}(v,T=0)$ as the coupling $\alpha$ becomes stronger. 
In Fig.~\ref{fig:PvT25}(d) the driving velocity is rather ``large'' ($\hbar v/\Delta^2=2$): here we observe that while for small
couplings ${\mathrm P}_{\gs}(t)$ is basically on top of the corresponding coherent evolution dynamics (including almost
invisible, on the scale of the figure, coherent dynamics oscillations), at larger couplings the drop of ${\mathrm P}_{\gs}(t)$
is rather sharp and non-oscillatory, again with a partial recovery which tends to push ${\mathrm P}_{\gs}(t_{\rm f})$
towards $P_{\gs}(v,T=0)$ as the coupling $\alpha$ becomes stronger.

The previous discussion can be fairly well summarized by saying that 
a ``thermally assisted'' QA dynamics requires necessarily some form of transverse noise, which we now explore in the
regime of finite bath temperatures $T>0$. 
The crucial question is: will a finite-$T$ bath improve or be detrimental to the $T=0$ dynamics in presence of a transverse noise?
We will see that the answer to such question depends on the driving velocity $v$. 
Anticipating the following discussion, we might say that increasing $T$ is helpful in the fast driving regime 
$\hbar v/\Delta^2\gg 1$, while it is rather detrimental in the ``adiabatic'' regime $\hbar v/\Delta^2\ll 1$.

Unfortunately, the QUAPI does not provide a very good benchmark for the $\theta=\pi/2$ case at finite temperature. 
The reason for this has to do with the Trotter error,  which is proportional to $|\epsilon(t)|(\delta t)^3$,  
as one can check by calculating the relevant commutator $[\hat{H}_{\rm QB} ,[\hat{H}_{\rm Q}(t),\hat{H}_{\rm QB}]](\delta t)^3$, 
where $\hat{H}_{\rm QB}$ is the system-bath interaction. 
This implies that the Trotter error is very large at the initial and final stages of the evolution, where $|\epsilon(t)|=v|t|\gg \Delta$
unless the corresponding Trotter time $\delta t$ is decreased accordingly.
In practice, this makes large values of the evolution time-interval $[-t_a,t_a]$ intractable with QUAPI.
Nevertheless, we have benchmarked our finite-$T$ QME data by comparing against QUAPI the results of evolutions restricted to smaller
time-intervals, with $vt_a\approx 20\Delta$, for which however ${\mathrm P}_{\gs}(t_{\rm f})$ deviates quantitatively from $P_{\rm gs}(v,T)$.
In this way we verified a rather good agreement (not shown) between QUAPI and QME also for transverse noise, for all the temperatures 
we have studied, at least in the weak-coupling region.  
Armed with this satisfactory validation of our weak-coupling finite-temperature QME approach, 
we go on and we consider a transverse noise at finite temperature in the long-evolution time regime, $vt_a \gg \Delta$.  
Here the data show a very intriguing behavior with temperature, as summarized in Fig.~\ref{fig:summary}(b). Remarkably, for a sufficiently fast driving $\hbar v/\Delta^2\gg 1$, a higher bath temperature can significantly enhance the performance of the annealing protocol with respect to the $T=0$ case, hence effectively providing a ``thermally assisted'' QA \cite{Amin_PRL08}.
On the contrary, this beneficial effect of a bath-temperature increase disappears, turning into detrimental, in the opposite regime of 
small driving velocity $\hbar v/\Delta^2\ll 1$.

\begin{figure}
\begin{center}
  \includegraphics[width=\columnwidth]{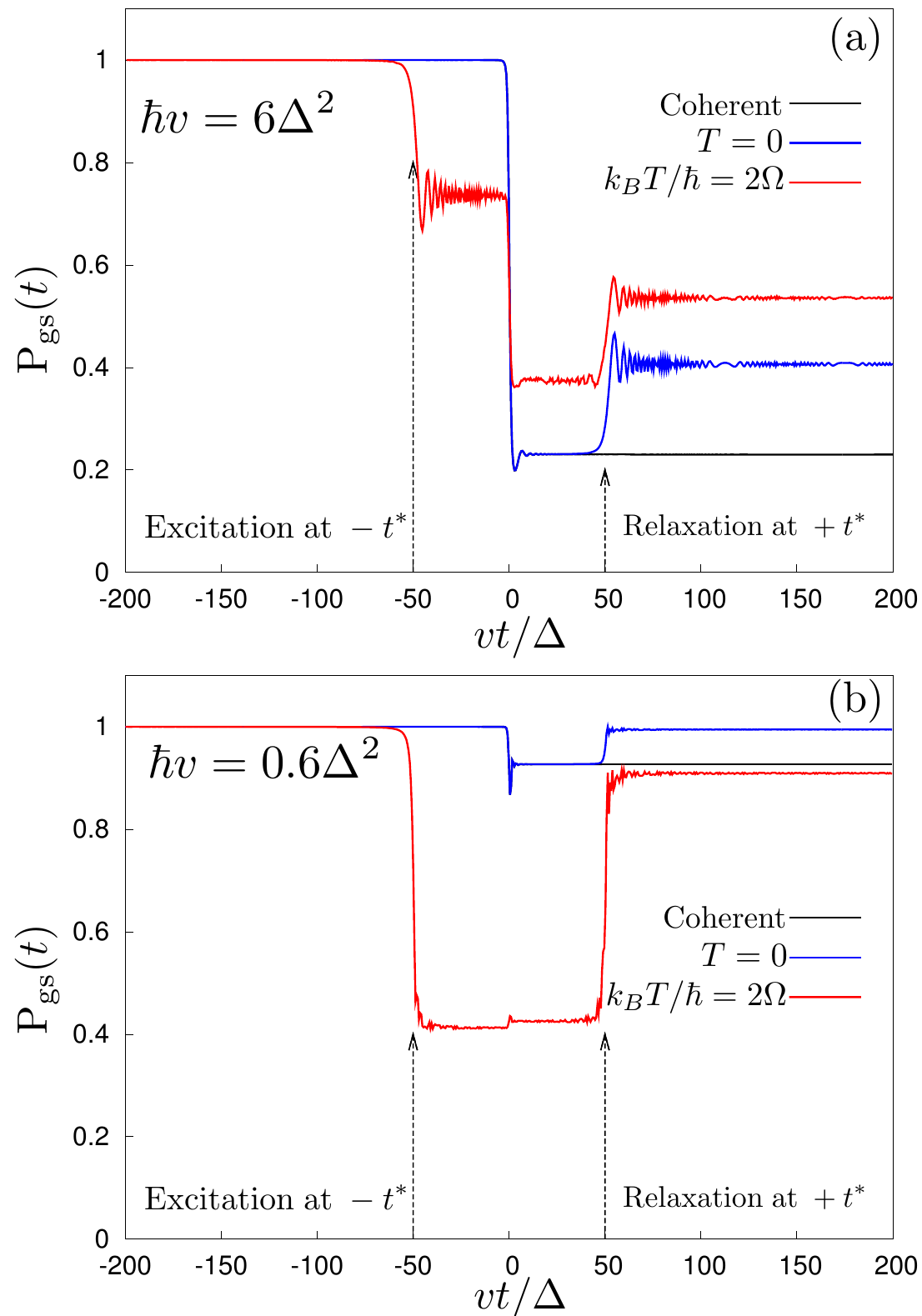} 
\end{center}
\caption{Single oscillator approach. Ground state probability ${\mathrm P}_{\gs}(t)$ {\it versus} time for (a) $\hbar v/\Delta^2=6$ and 
(b) $\hbar v/\Delta^2=0.6$. Here $\lambda=0.5 \Delta$ and $\theta=\pi/2$.  
The black line is the coherent evolution ${\mathrm P}_{\gs}(t)$ in absence of dissipation; 
the blue and red lines are ${\mathrm P}_{\gs}(t)$ at $T=0$ and $k_BT=2\hbar \Omega$, respectively. 
In both cases $\hbar\Omega=50\Delta$. 
Vertical lines at $\pm t^*$, the resonance times in Eq.\eqref{resonance_time}, highlight the excitation/relaxation mechanisms. 
}
\label{fig:SingleOscillator}
\end{figure}

To better understand the physics behind this effect, we consider a drastically simplified model of ``environment'', which is reduced
to a {\em single harmonic oscillator} \cite{Ashhab_PRA14}, with a fixed frequency $\Omega$ coupled to the qubit along a fixed direction in spin-space, 
parameterized as before with the angle $\theta$.
The Hamiltonian is therefore:
\begin{equation}	\label{ham_sing_osc}
\hat{H}(t)= \hat{H}_{\rm Q}(t) + \hbar \Omega \opbdag{} \opb{} \!+ 
\lambda (\cos{\theta} \PauliSigma^z + \sin{\theta} \PauliSigma^x) (\opbdag{} \!+ \opb{}\!\!)  \;.
\end{equation}
According to the predictions of Ref.~\onlinecite{Saito_07}, such a zero-temperature ``single-oscillator bath'' would still 
lead to a $P_{\gs}(v,T=0)$ given by Eq.~\eqref{T0_Saito_formula}, where now:
\begin{equation} \label{eqn:single_W}
	W_{\theta}^2 = \Big| \Delta - \frac{2\lambda^2}{\hbar\Omega} \sin{2\theta} \Big|^2 + 4\lambda^2 \sin^2\theta \;.
\end{equation}
Notice that we can tune $\Omega$ and $\lambda$ so as to get the same $W_{\theta}^2$ we would have for a set of infinitely many 
harmonic oscillators with Ohmic spectrum (see Eq. (20) in Ref.~\onlinecite{Saito_07}).
Therefore, the behaviour of $P_{\gs}(v,T=0)$ for a Ohmic dissipative problem can be perfectly mapped into a specific 
``single-oscillator environment'' coupled to the system. 
This analogy, which is in principle justified only for $T=0$ and for the infinite-time Landau-Zener problem,
helps elucidating some of the physics of the problem, which becomes very transparent in the single-oscillator setting. 
In the following we will take $\hbar\Omega=50\Delta $ and a coupling $\lambda=0.5 \Delta$. 
The time evolution of this simple driven qubit can be studied both at zero and finite temperature by means of an exact 
diagonalization-based Schr\"odinger dynamics, provided the oscillator Hilbert space is properly truncated. 
Fig.~\ref{fig:SingleOscillator} shows the results of such a study for a purely transversal coupling $\theta=\pi/2$ in the non-adiabatic regime, 
$\hbar v/\Delta^2=6$ (a), and in the ``adiabatic'' regime $\hbar v/\Delta^2=0.6$ (b), compared to the ``free'' coherent evolution.
%
%
We observe that a $\PauliSigma^x$-coupling ``bath'' is indeed active when the instantaneous gap 
of the qubit $E(t)=\sqrt{(vt)^2+\Delta^2}$ matches exactly the oscillator energy $\hbar\Omega$. 
This resonance condition happens at two times $\pm t^*$ with 
\begin{equation} \label{resonance_time}
t^*= \frac{1}{v}\sqrt{(\hbar\Omega)^2-\Delta^2} \;, 
\end{equation}
which correspond to excitation (at $-t^*$, before the avoided crossing) and relaxation (at $+t^*$, after the avoided crossing) of the system, 
by absorption/emission of a quantum of vibration.
The final result is however very different depending on the driving velocity $v$.
In the fast-driving regime, Fig.~\ref{fig:SingleOscillator}(a), immediately after the avoided crossing ($0<t<+t^*$), 
the system ends up in the excited state with a significant probability, both at zero and at finite temperature. 
Thus, a relaxation mechanism occurring at $t=+t^*$ is quite effective in increasing ${\mathrm P}_{\gs}(t)$  
(the $\PauliSigma^x$-coupling providing the necessary matrix element at this resonant condition) 
up and above at the coherent-dynamics probability (solid black line).
We also observe that, at $T=0$ (blue solid line), the relaxation associated to a resonant emission of a quantum of vibration at $t=+t^*$
is the only possible process, since the oscillator is unable to excite the system at $t=-t^*$.
On the contrary, for $T>0$ (red solid line), the bath can excite the qubit {\em before} the avoided crossing, 
by a resonant absorption of a quantum of vibration at $t=-t^*$.
This effect, which reflects itself in a marked decrease of ${\mathrm P}_{\gs}(t)$ at $t\sim -t^*$, is actually beneficial to the
final ground state probability after the avoided crossing (where the instantaneous ground state is indeed flipped), 
providing an enhancement of ${\mathrm P}_{\gs}(t)$ over the coherent evolution result at times  $0<t<+t^*$
(red solid line). 
Finally, the subsequent relaxation process at $t=+t^*$, although less effective than at $T=0$, further improves the final probability
${\mathrm P}_{\gs}(t_{\rm f})$ up and above the coherent evolution result.
This state of matters changes if the driving velocity is small, as shown in Fig.~\ref{fig:SingleOscillator}(b),
corresponding to $\hbar v/\Delta^2=0.6$. Here, the coherent dynamics is nearly adiabatic. 
While a zero temperature transverse bath further improves $P_{\gs}(v,0)$ above $P_{\gs}^{\LZ}$, at finite $T$ the combined effect 
of the excitation at $t=-t^*$ and the subsequent relaxation at $t=+t^*$ is eventually slightly detrimental.

\section{Discussion and conclusions} \label{sec:conclusions}
%
In this study, we have investigated the role of the bath temperature and of the spin-coupling direction in a simple dissipative LZ model, showing that thermally assisted AQC requires a transverse component of the coupling and is generally effective only in the fast driving regime.

We can now return to the results of Ref.~\onlinecite{Dickson_NatCom13} and discuss them in the light of our findings.
Ref.~\onlinecite{Dickson_NatCom13} deals with an explicit realization of an approximate two-level system LZ dynamics 
using 16-qubit of the D-Wave$^{\mbox{\tiny \textregistered}}$ machine.
The gap at the anti-crossing between the two low-lying instantaneous eigenstates is engineered to be rather small,
$\Delta=g_{\rm min}=0.011$ mK$/k_B$, compared to a rather large energy separation, $\delta E\approx 50.5$ mK$/k_B$, from 
all the higher-lying states.
In the experiment, several annealing runs are performed at different driving velocities and temperatures, with $T$
ranging from $T_{\rm low}=19.9$ mK up to $100.8$ mK. Particularly interesting are the first four temperature datasets, 
ranging from $T_{\rm low}$ up to $T_{\rm high}=34.9$ mK, where the two-level-system approximation is reasonable,
since $k_B T< \delta E$. 
The extrapolation of the lowest experimental annealing data down to $T=0$ (see Fig.~3 of Ref.~\onlinecite{Dickson_NatCom13}) 
allows to extract $P_{\rm gs}(v,T=0)$, which closely matches a Landau-Zener type of expression.
This is, in a sense, not surprising, in view of the exact results of Refs.~\onlinecite{Wubs_06,Saito_07}; it is also perfectly
in line with the Fermi golden rule findings in the incoherent-tunnelling-dominated regime associated to a significant sub-ohmic 
flux noise \cite{Amin_MRT_PRL08}, as well as with the seminal analysis by Ao and Rammer \cite{AoRammer_PRB91}.
But, as we have discussed, a LZ form is rather generic: it does not tell you if the coupling is predominantly in the longitudinal
$\PauliSigma^z$-direction, or if there is some component of transverse noise. 
The finite-$T$ experimental curves with $T=19.9 \div 34.9$ mK are in some sense a strong argument in favour of the fact
that  {\em there should be some transverse noise} affecting the annealing dynamics. 
Indeed, Fig.~4 of Ref.~\onlinecite{Dickson_NatCom13} clearly shows that $P_{\rm gs}(v,T)$ is considerably improved
over the $T=0$ curve in its fast-driving tail, which would be impossible with a purely longitudinal noise.
Consistently with our findings, this ``thermally assisted'' QA turns into a detrimental effect in the slow-driving regime.
What is hard to explain from our very rough modelling is the fact that the experimentally extracted $P_{\rm gs}(v,T)$ 
remains quite different from $1/2$ --- the value you would expect at $T=\infty$ --- at temperatures which are incredibly large 
compared to the minimum gap, $k_BT/\Delta = 1809 \div 3172$.
Evidently, a detailed understanding of the experimental findings needs a more refined modelling, possibly including the 
unavoidable time-dependence of the couplings $g_i(t)$ to the environment, as well as the possible presence of sub-ohmic noise
components, very hard to tackle with traditional QME weak-coupling techniques. 

In future, the role of a $\PauliSigma^y$-coupling, which we have not explicitly addressed, might also be worth looking at.
Indeed $\PauliSigma^y$ is precisely the Hamiltonian term realising the {\em shortcut to adiabaticity} \cite{Demirplak_JPCA03},
or {\em transitionless quantum driving}, in Berry's terminology \cite{Berry_JPA09}, in our LZ problem: this is clear from
the  presence of the $\dot{\phi}_t \, \PauliSigma^y$ term in the rotated Hamiltonian in Eq.~\eqref{RotatedH:eqn}, 
see also Ref.~\onlinecite{delCampo_PRL12}. 
Clearly, the possible time-dependence of the bath-couplings, inherited by projecting the two lowest-lying instantaneous 
eigenstates into an effective two-level system, might also play a role, especially in view of the fact that the larger transverse 
field present before the anti-crossing might favour thermal excitations over the thermal relaxations phenomena
occurring after the anti-crossing \cite{Dickson_NatCom13}. 
Further work is necessary to fully elucidate all these aspects.

\section*{ACKNOWLEDGMENTS}
We acknowledge fruitful discussions with S. Suzuki, M. Keck and D. Rossini.
RF kindly acknowledges support from the National Research Foundation of Singapore (CRP - QSYNC) and the Oxford Martin School.
Research was partly supported by EU ERC MODPHYSFRICT.

\vspace{3mm}

\appendix
\section{A few remarks on the derivation of the quantum master equation}  \label{appA}
In the usual setting of a perturbative QME approach, the bath has a large number of degrees of freedom weakly coupled to the system, which in turn perturbs the bath state in a negligible way.   
Hence, the bath $\hat{X}$-correlation function is essentially unmodified by the system, and is given by 
\begin{align} \label{eqn:Ct}
C(t) &=\Tr_{\rm B} \left( \hat \rho_{\rm B} \hat{X}(t) \hat{X}(0) \right) =  \nonumber\\
	&=\int_0^{\infty} \! d\omega \; J(\omega) \Big(\nep^{i\omega t} f_{\rm B}(\omega) + \nep^{-i \omega t} [f_{\rm B} (\omega)+1] \Big) 
\end{align}
where $\hat{X}(t)=\nep^{i\hat{H}_{\rm B}t/\hbar} \hat{X} \nep^{-i\hat{H}_{\rm B}t/\hbar}$ (in the usual interaction representation) 
and the second expression applies to a strictly harmonic bath for which 
$\Tr_{\rm B} \left( \rho_{\rm B} \opbdag{k}\opb{k}\right) = f_{\rm B}(\omega_k)= 1/(\nep^{\beta \hbar \omega_k} -1)$ 
is the Bose distribution. 
Ideally, for a Markovian approximation to hold, one would require that $C(t)$ decays fast with respect to the timescales 
of the evolving system; this is strictly speaking not the case for an ohmic bath $C(t)$, which shows a power-law tail even at finite temperature. 
We consider the case of a system weakly coupled to its environment, under the usual Born-Markov approximations \cite{Cohen:book,Gaspard_JCP99a,Yamaguchi_PRE17}.
Following Ref.~\onlinecite{Nalbach_PRA14}, we perform a time-dependent rotation in spin space around the $y$-axis, 
$\hat{R}_t = \exp[i\phi_t\PauliSigma^y/2]$, with $\phi_t = \arctan(\epsilon(t)/\Delta)$, such that
$\hat{R}_t^\dagger \hat{H}_{\rm Q}(t) \hat{R}_t = -E_t \PauliSigma^x / 2$, 
with $E_t \equiv \hbar\Lambda_t \equiv \sqrt{\Delta^2 +\epsilon^2(t)}$.
This leads to an effective qubit Hamiltonian of the form: 
\begin{equation} \label{RotatedH:eqn}
{\tilde{H}}_{\rm Q} (t) = \hat{R}_t^\dagger \hat{H}_{\rm \rm Q}(t) \hat{R}_t + i\hbar \frac{d \hat{R}^{\dagger}_t}{dt} \hat{R}_t                                   
                           = -\frac{E_t}{2} \PauliSigma^x + \frac{\hbar\dot{\phi}_t}{2} \PauliSigma^y \;.
\end{equation}
In this time-dependent frame we denote by a tilde the other rotated operators, 
$\tilde{\rho}_{\rm Q}(t) = \hat{R}_t^\dagger \hat{\rho}_{\rm Q}(t) \hat{R}_t$ and 
$\tilde{\sigma}_\nu(t) = \hat{R}_t^\dagger \PauliSigma^\nu \hat{R}_t$, and write the Schr\"odinger picture QME as:
\begin{equation} \label{QME_rotated}	
	 \frac{d \tilde{\rho}_{\rm Q}}{dt} = -\frac{i}{\hbar} \left[ \tilde{H}_{\rm Q}, \tilde{\rho}_{\rm Q} \right] 
- \frac{1}{\hbar^2} \sum_{\nu,\nu'}^{x,z} g_\nu^2 \left( \left[ \tilde{\sigma}_\nu, \tilde{S}_{\nu'} \tilde{\rho}_{\rm Q} \right] + \hc \right) \;.
\end{equation}
The effect of the bath is all encoded in
\begin{equation}  
	\tilde{S}_{\nu}(t) =\int_{0}^{t-t_0} \hspace{-6mm} d\tau \, C(\tau) \, {\tilde{\mathcal U}}_{\rm Q}(t,t-\tau) \tilde{\sigma}_{\nu}(t-\tau) 
\mathcal{\tilde U}^{\dagger}_{\rm Q}(t,t-\tau) 
\end{equation}
where
\begin{equation} 
\mathcal{\tilde U}_{\rm Q}(t,t-\tau)=\Texp \left[ - \frac{i}{\hbar}\int_{t-\tau}^t \hspace{-3mm} dt' \, \tilde{H}_{\rm Q}(t') \right] \equiv 
\nep^{ -\frac{i}{\hbar} \overline{\rm H} \, \tau } 
\end{equation}
is the free evolution operator associated to $\tilde{H}_{\rm Q}(t)$, which we have re-expressed 
in terms of an appropriate effective Hamiltonian $\overline{\rm H}$, in general function of both $t$ and $\tau$.  
In the well-known setting of a time-independent qubit Hamiltonian $\hat{H}_{\rm Q}$, we would have $\overline{\rm H}=\hat{H}_{\rm Q}$ and one would then argue that, assuming $C(\tau)$ to decay on a fast enough timescale $t_{\rm B}$, 
the upper limit of the integral in $\tilde{S}_{\nu}$ can be effectively
set to infinity \cite{Gaspard_JCP99a,Yamaguchi_PRE17} (as soon as $t-t_0\gg t_{\rm B}$),
and any other dependence on $\tau$, apart from that of the free-evolution operator, can be neglected.
In the present time-dependent framework, these two standard approximations lead to:
\begin{equation}
	\tilde{S}_\nu(t) \approx \int_{0}^{\infty} \hspace{-2mm} d\tau \, C(\tau) \,
\nep^{-\frac{i}{\hbar} \overline{\rm H} \, \tau} \, \tilde{\sigma}_{\nu}(t) \, \nep^{\frac{i}{\hbar} \overline{\rm H}\, \tau} 
\;. 
\end{equation} 
A further approximation that we do is to take the evolution operator effective Hamiltonian $\overline{\rm H}$
to be simply $\overline{\rm H} \approx-E_t\PauliSigma^x/2=\hat{R}_t^\dagger \hat{H}_{\rm Q}(t) \hat{R}_t$.
Although not straightforward to justify, especially in the non-adiabatic large-$v$ regime, we will assume it here, 
following the literature \cite{Nalbach_PRA14}, and refer to Ref.~\onlinecite{Yamaguchi_PRE17} for a detailed discussion
of the different time-scales involved in the dynamics.
The great advantage of this approximation is that it allows us to analytically calculate the rate constants
appearing in our QME: indeed, the effect of dissipation reduces to that of an unbiased spin-boson problem with an 
{\em instantaneous} value of the tunnelling amplitude $\Delta \mapsto E_t=\hbar\Lambda_t$. 
We will test the approximations involved in this QME {\em a posteriori}, against numerically ``exact'' QUAPI data and 
exact zero temperature analytical results \cite{Wubs_06,Saito_07}. 
Eventually, neglecting the so-called {\em Lamb shift}\cite{Cohen:book} terms due to the small imaginary parts in the Fourier transform of 
$C(t)$ and by adopting the Bloch-sphere representation 
$\tilde{\rho}_{\rm Q}(t) = \frac{1}{2} (\id + \sum_{\nu}r_{\nu}(t)  \PauliSigma^{\nu})$ with $\nu=x,y,z$,
we obtain the weak-coupling QME shown in Eq.\eqref{QME_noRWA}.
\begin{figure}[b]
\begin{center}
  \includegraphics[width=\columnwidth]{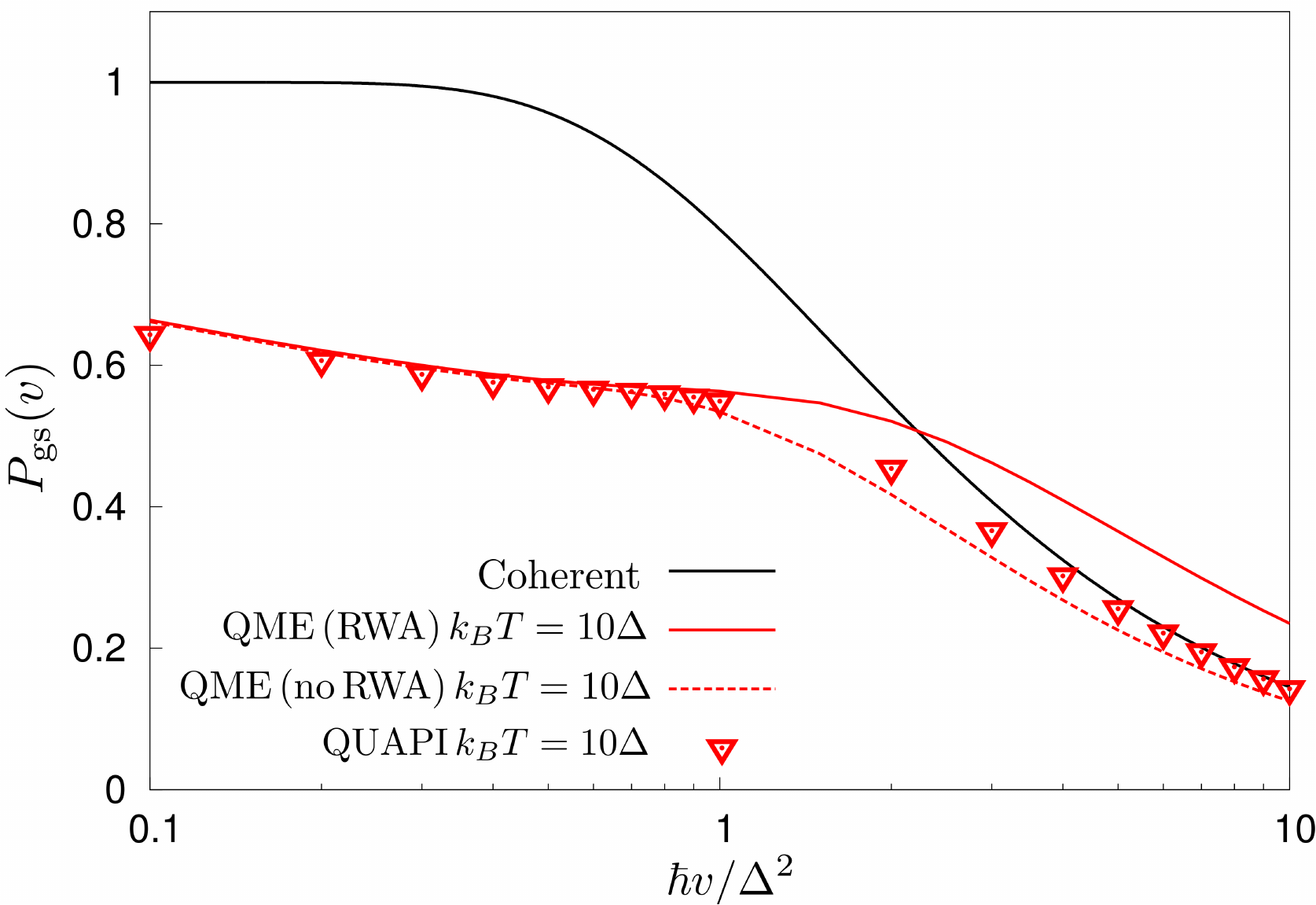} 
\end{center}
\caption{Comparison of QME results obtained with and without the RWA for a longitudinal bath coupling with
$\alpha=0.02$ and $k_BT = 10 \Delta$. The points denote numerically exact QUAPI data. The RWA line shows
a quite clear over-shooting above the coherent evolution result which is an artefact of the approximation. 
}
\label{fig:RWA}
\end{figure}
\section{Rotating-wave approximation} \label{appB}
The {\em rotating wave approximation} (RWA) \cite{Cohen:book} is a very popular approximation which, among other things, allows
to put the QME, in the time-independent case, into an explicitly Lindbladian form. 
This, in particular, guarantees that the approximation explicitly preserves the positivity condition of the density matrix.
There is no really compelling reason to adopt it in the present time-dependent case, apart from a small simplification
of the QME equations, which will now read as follows \cite{Nalbach_PRA14,Javanbakht_15}:
\begin{equation} \label{QME_RWA}
\left\{ \begin{array}{lcl}
	\dot{r}_x &=& -\gamma_{x} (r_x - \overline{r}_x) + \dot{\phi}_t r_z \\
	\dot{r}_y &=& -\gamma_{\Deco} r_y + \Lambda_t r_z \\
	\dot{r}_z &=& -\dot{\phi}_t r_x  - \Lambda_t r_y -\displaystyle \gamma_{\Deco}  \, r_z \;.
\end{array}
\right. 
\end{equation}
Overall, we find that the difference in the results obtained with and without RWA is rather small if one stays
in the weak-coupling region. 
In particular, we have verified that our density matrix remains positive definite at all times, even if we do not explicitly use a RWA. 
Nevertheless, the RWA results tend to produce, as an artefact of the approximation, a certain tendency
to increase $P_{\gs}(v,T)$ in the large $v$ tails, up and above the coherent evolution result.  
We show this in Fig.~\ref{fig:RWA}, where we plot QME results obtained with and without RWA ({\it i.e.}, using
the full set of Eqs.\eqref{QME_noRWA}) benchmarked against numerically exact QUAPI data at the
moderate coupling $\alpha=0.02$. 
The RWA line shows a quite clear over-shooting above the coherent evolution result which is an artefact of the approximation.


\end{document}